# Measurement of resistance coefficients of pendulum motion with balls of various sizes

Kyung-Ryul Lee and Young-Gu Ju*

Department of Physics Education, Kyungpook National University, Daegu 41566

In order to obtain the damping and resistance coefficients of a pendulum, we constructed an optical system containing a photogate for measuring the speed of the pendulum at the lowest point of motion. The photogate consisted of a photoresistor, a laser, a mechanical body, and a pendulum ball. A 3D printer was used to produce the mechanical body and pendulum balls of various sizes. Furthermore, we used Arduino to automate measurement of the speed at the lowest point of motion and increase the precision. We found that the resistance coefficient was proportional to the size of the balls, regardless of the ball mass, in agreement with the drag equation for a small Reynolds number.

# I. INTRODUCTION

A pendulum motion is one of the most common topics in physics textbooks, especially for explaining Newtonian mechanics, including the concepts of periodic motion, gravity, oscillation, and friction.[1-3] For small-angle oscillations, the pendulum motion reduces to a harmonic oscillator problem. Many theoretical and experimental studies related to simple pendulums have been conducted to investigate its nonlinear effects and improve the precision of measurements.[4-6]

The pendulum motion can be described by a sinusoidal function of time if there is no air resistance. The frictional force associated with air resistance causes damping of the pendulum motion, which can be described by the same equation used for a damped oscillator. Since the air resistance is generally not sufficiently large to cause overdamping or critical damping, a damped pendulum can be considered as an underdamped oscillator, which has an amplitude that decays exponentially in time with a damping coefficient.[1,2] The damping coefficient is equal to the resistance coefficient of the friction term in the equation of motion, divided by two times the mass. In most textbooks, this friction term is assumed to be the product of the resistance coefficient and the velocity of the bob or ball of the pendulum. Since air friction is a type of drag force, it should depend on the size of the pendulum ball. It is interesting to investigate the dependence of the resistance coefficient on the size of the pendulum ball.

In this study, we measured the damping and resistance coefficients for pendulum balls of various sizes. We examined the dependence of the resistance coefficient on the ball size and examined the compatibility of the results with the drag equation, which is frequently used to describe the air resistance acting on the motion of spherical objects.[7,8]

Another salient feature of this research is the achievement of high accuracy in and simplification of the pendulum experiment for educational purposes. We employed a microcontroller-based laboratory method[9] and used Arduino,[10] which is one of the most popular open-source electronics platform that offers easy-to-use hardware and software. This electronics platform has considerable advantages in terms of cost, availability, and ease of learning,[11,12] which are especially helpful for teaching students basic experiments such as an experiment for understanding pendulum motion. Arduino can be used to automate data acquisition and increase the measurement accuracy by sampling more data with ease.

In addition to Arduino, we employed a 3D printer to fabricate the mechanical structure of a photogate and balls of various sizes. A 3D printer are cheap and easily available in secondary schools. The combination of Arduino and a 3D printer can help students perform many basic experiments without having to purchase expensive instruments, and provide students with an opportunity to learn coding.

## II. THEORY

According to fluid dynamics, the frictional force (or drag) for a spherical object in motion depends on the properties of fluid and on the size, shape, and speed of the object. This relationship is described by the drag equation, which is expressed as Eq. (1), where $F_d$, $C_D$, $\rho$, $A$, and $v$ are the drag force, drag coefficient, density of the fluid, cross-sectional area, and velocity of the object relative to the fluid, respectively.[7,8]

$$F_d = \frac{1}{2} C_D \rho A v^2 \qquad (1)$$

In addition, $C_D$ depends on the Reynolds number, which can be derived using Eq. (2), where L is a characteristic diameter or linear dimension of the object such as the diameter of a sphere and µ is the dynamic viscosity of the fluid.

$$R_e = \frac{\rho L |v|}{\mu} \qquad (2)$$

If the Reynolds number $R_e$ is about 1200 or less, the drag coefficient $C_D$ is roughly inversely proportional to it. [7,13] This means that the drag force is proportional to the velocity as in Eq. (3), where c is the resistance coefficient.:

$$F_d = -cv \qquad (3)$$

Therefore, for a pendulum motion moving with a small Reynolds number, the frictional force $F_d$ due to air resistance is proportional to the velocity of the object as shown in Eq. (3) where the resistance coefficient *c* is independent of the velocity, but is proportional to the size of a sphere. In our experiment, the measured $R_e$ was between 296 and 927, which is less than 1200: the drag was thus considered proportional to the velocity as in Eq. (3).

Using the drag equation, the following differential equation for pendulum motion can be obtained:

$$ml\ddot{\theta} + cl\dot{\theta} + mg\sin\theta = 0, \qquad (4)$$

where m, c, g, and l are the mass, resistance coefficient, standard gravity, and pendulum length, respectively.[1,2] For a small-angle oscillation, this expression can be approximated as

$$m\ddot{\theta} + c\dot{\theta} + (mg/l)\theta = 0. \tag{5}$$

The solution of this differential equation has the form of Eq. (6), where $\theta_m$, $\gamma$, and $\omega$ are the maximum angle, damping coefficient, and angular frequency, respectively. Furthermore, the angular frequency can be expressed in terms of the standard gravity and pendulum length, as shown in Eq. (7). The damping coefficient is related to the resistance coefficient according to Eq. (8):

$$\theta(t) = \theta_m e^{-\gamma t} \cos(\omega t) \tag{6}$$

$$\omega = \sqrt{\frac{g}{l}} \tag{7}$$

$$\gamma = \frac{c}{2m} \tag{8}$$

The expression for the angular velocity of the pendulum ball [Eq. (9)] can be obtained by differentiating Eq. (6) with respect to time:

$$\dot{\theta}(t) = \theta_m(-\gamma e^{-\gamma t}\cos(\omega t) - \omega e^{-\gamma t}\sin(\omega t)). \tag{9}$$

At the lowest point of motion, $cos(\omega t)$ becomes zero and $sin(\omega t)$ becomes ±1, and the following expression can be obtained:

$$\theta(t_{low}) = 0 = \theta_m e^{-\gamma t_{low}}\cos(\omega t_{low}). \tag{10}$$

Substituting Eq. (10) in Eq. (9) gives

$$\dot{\theta}(t_{low}) = \theta_m(0 - \omega e^{-\gamma t_{low}}\sin(\omega t_{low})) = \pm\theta_m\omega e^{-\gamma t_{low}} = \pm\theta_m\omega e^{-\gamma t_{low}} \quad . \qquad (11)$$

By multiplying both sides of Eq. (11) by $l$, we can obtain the following expressions for the linear velocity and pendulum speed at the lowest point:

$$v(t_{low}) = \pm v_m e^{-\gamma t_{low}}, \qquad (12)$$

$$|v(t_{low})| = v_m e^{-\gamma t_{low}}, \qquad (13)$$

$$v_m = \theta_m l\omega. \qquad (14)$$

The speed at the lowest point is also the maximum speed of the pendulum in any given period of motion. The speed of the pendulum decreases exponentially with time, and the decay constant equals the damping constant of the motion. Taking the natural logarithm of both sides of Eq. (13) gives

$$ln|v(t_{low})| = -\gamma t_{low} + ln(v_m). \qquad (15)$$

When $ln|v(t_{low})|$ is plotted as a function of $t_{low}$, the slope of the line corresponds to the damping coefficient with the opposite sign. Thus, the damping coefficients of pendulum motion can be obtained from the measured speed of the ball at the lowest point.

## III. EXPERIMENTS

In order to measure the speed of a pendulum ball at its lowest point, we fabricated a photogate using Arduino and a 3D printer in the laboratory. The photogate consisted of a diode laser, a photoresistor, a mechanical structure for holding the two devices, and an Arduino circuit, as shown in Fig. 1. We used Cubify Invent as the computer-aided design (CAD) software to design the photogate body and pendulum balls. The photogate body was U-shaped with two holes at the ends [Fig. 1(a)] for holding the diode laser and photoresistor. The pendulum ball swung between these two devices blocking the laser beam and decreasing the laser light arriving at the photoresistor.

The photoresistor formed the upper part of a voltage divider with a fixed resistor on a breadboard. The upper and lower ends of the voltage divider were connected to the 5 V pin and 0 V (GND) pin of an Arduino board, respectively, as shown in Fig. 1(b) and (c). The center of the voltage divider was tied to an analog input labeled "A2" on the Arduino board. When the laser beam was not blocked, the light reached the photoresistor and decreased its resistance, thereby reducing the voltage drop across the photoresistor; this caused the voltage at the middle point of the voltage divider to remain high. As the ball began to block the laser beam, the process was reversed: the voltage at the middle point and A2 pin of the Arduino board decreased. In short, a lower light intensity at the sensor resulted in a lower voltage at the Arduino input and vice versa. Thus, the voltage of the photosensor indicated whether the ball was blocking the laser beam.

Since the Arduino board could check the sensor very frequently and measure time with an accuracy of a tenth of a millisecond, we could accurately measure the time span between the instant when the ball began blocking the laser beam and when the blocking ended (i.e., when the pendulum was at its lowest point). The speed of the ball was the diameter of the ball divided by the blockage time span. The experimental setup is shown in Fig. 2. The ball was suspended by two strings instead of a single string to prevent the rotation of the plane of the pendulum motion during the measurements.

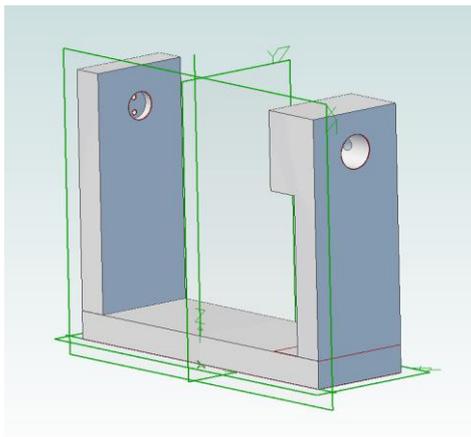

(a)

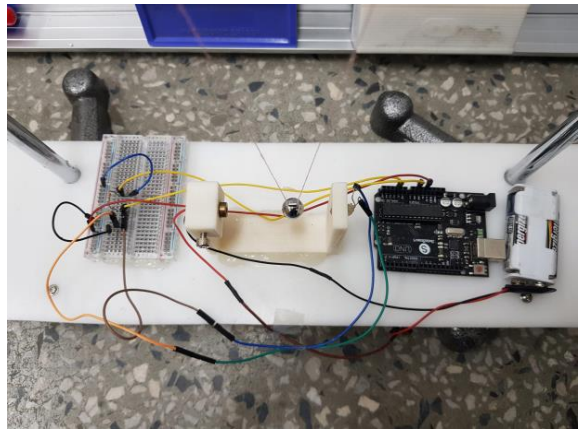

(b)

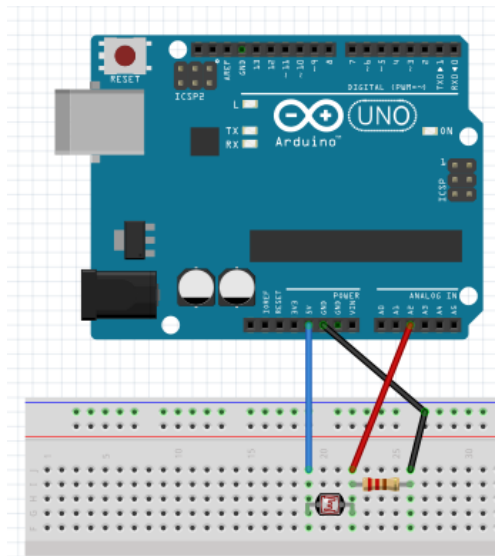

(c)

FIG. 1. Photogate fabricated using the 3D printer and Arduino: (a) a schematic of an assembly

view of the photogate body and photographs of (b) the photogate comprising the laser, photoresistor, and Arduino circuit and (c) the detailed wiring of the divider with the photoresistor and a resistor connected to the Arduino board.

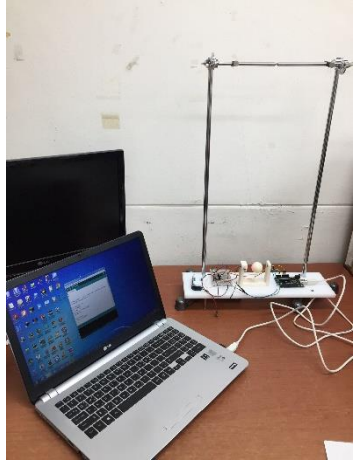

FIG. 2. Photograph of the setup for the simple pendulum experiment.

Pendulum balls of various sizes were designed and fabricated using CAD and the 3D printer, and they are shown in Fig. 3. Basically, a single spherical metal ball, shown in Fig. 3(c), was used throughout the experiment with strings of a fixed length. The ball was enclosed by plastic shells of different sizes to vary the bob size. This approach eliminated possible changes in the experimental conditions such as the length of strings. The plastic shell consisted of two halves that could be attached to each other with two pins. A half-shell had two holes on the flat side for holding the pins, as shown in Fig. 3(a). The plastic shell also had space at its top for two strings to pass through. The outer diameters of the spherical shells were 18.7, 24.1, 30.0, and 36.0 mm. Balls with five different sizes, including the metal ball (which had a diameter of 11.5 mm), were used in the pendulum experiment.

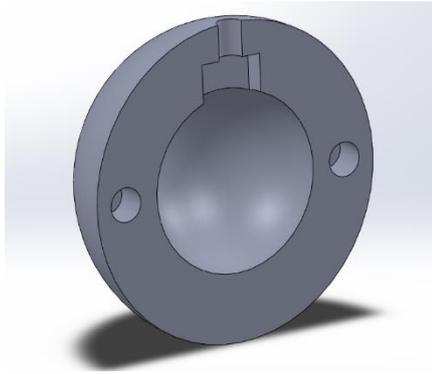
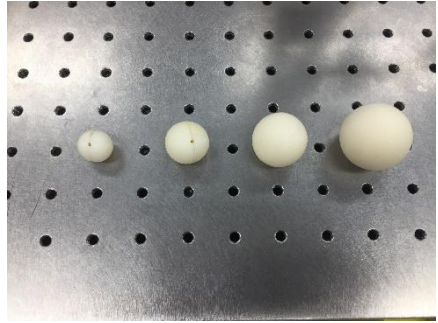

(a) (b)

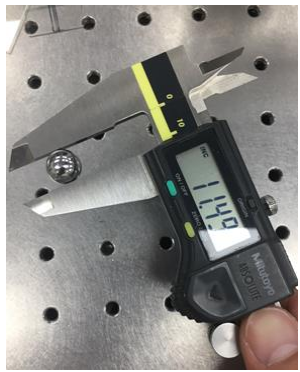

(c)

FIG. 3. Balls used in the experiment

(a) the 3D design of a plastic ball and photographs of (b) plastic balls of various sizes produced by the 3D printer and (c) the metal ball that was inserted into the plastic shells.

The measurement of pendulum motion required the programming of the Arduino board to acquire time data at the exact instants when the ball began to block the laser path and when the laser path was unblocked. The program used in the experiment is shown in Fig. 4. It enabled the board to check whether the voltage at the A2 pin was high or low. If the voltage was greater than 2 V, then the program recognized the state as HIGH. Otherwise, the state was identified as LOW. The HIGH and LOW states corresponded to high and low light intensities at the photoresistor, respectively. Since the voltage input at the A2 pin was digitized into an integer value from 0 to 1023, the threshold value in the code was set to 400. The measured timing data were stored in the memory until the number of data reached 100, whereupon the board sent the stored data in a single batch through serial communication to the main computer to prevent unnecessary and irregular latency in the communication between data acquisition intervals.

The clock frequency of the Arduino board was 16 MHz, and the number of coding lines in the execution loop was less than 10. Therefore, theoretically, the time taken to check the photosensor voltage should be less than a microsecond. However, the time to execute a loop was measured to be about 130 µs. It emerged that reading the voltage from an analog input was relatively slower than the execution of the other instructions and consumed most of the execution time. Since the laser path was blocked for a time span of a few tens of milliseconds, the relative error in the time measurement was on the order of 1/100 or less. The relative error in the measurement of the ball diameter was on the order of 1/1000 or less since the diameter was about a few tens of millimeters and the precision of the vernier calipers was 0.01 mm. Therefore, the precision of the time span was critical to the measurement precision for the speed of the ball.

```
photo_gate2 §
int photocellPin = 2; //define photocellsh=2, read the value of voltage.
int val = 0; //define original of val.
unsigned long time1=0;
unsigned long time2;
unsigned long time3;
unsigned long dt=0;
int button_old = LOW;
int debounce_limit = 100;
int buttonState;
int count = 0;
void setup() {
  Serial.begin(9600);
}
void loop() {
  val = analogRead(photocellPin); //get the value from sensor
  if( val < 400)
  {
    buttonState = LOW;
  }
  else{
    buttonState = HIGH;
  }
  if(button_old ==  LOW && buttonState == HIGH)
  
    time2= millis();
    dt = time2 - time1;
    if(dt > debounce_limit )
    {
      if(count%2 ==0)
      {
        if(count != 0)
          Serial.println(dt);
        time1 = time2;    // time reset
      }
      count = count+1;
    }
  }
  delay(1); // delay in between reads for stability
  button_old = buttonState;
}
```

FIG. 4. Arduino program used for time measurement by the photogate.

The experiment was performed as follows. First, the ball was pulled to a distance of 10 cm. After the activation of the Arduino board, the ball was released and allowed to oscillate. After the pendulum passed the optical sensor 100 times, Arduino sent the data to a notebook that displayed the information on its serial monitor. The data from the serial monitor was transferred to a spreadsheet program for additional analysis. The lengths of the strings were 50 cm. The experiment was conducted five times under the same conditions for every ball. In order to reduce the error, we took adequate care to ensure that there was no initial speed at the specified width of swing when releasing the pendulum ball. We used a ruler to stop a ball before commencement of its motion. Since the ruler was fastened by a screw, loosening the screw allowed the ball to move. This method helped initiate the motion under the same conditions in terms of the width of swing and initial speed.

## IV. RESULTS AND DISCUSSION

Using the fabricated photogate, we measured the speeds of the balls of various sizes during pendulum motion. In the experiment, the change in the speed of the oscillating bob with time was measured. From the rate of reduction of the speed, the damping and resistance coefficients could be determined. A change in the ball diameter resulted in changes in the resistance and damping coefficients, which were obtained from the speed of the pendulum at its lowest point. From plots of $ln|v(t_{low})|$ as a function of time, the slope and damping coefficient were obtained using Eq. (15). Plots of $ln|v(t_{low})|$ vs. time for the various ball sizes are shown in Fig. 5.

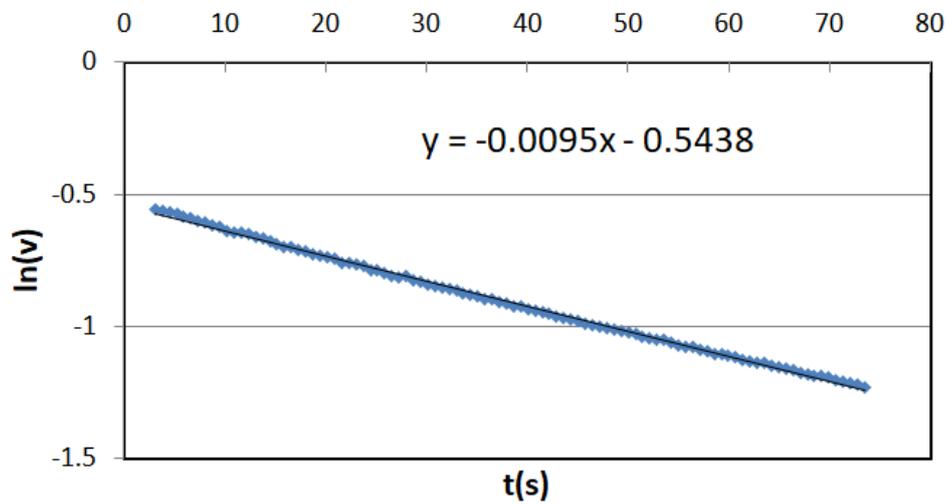

(a)

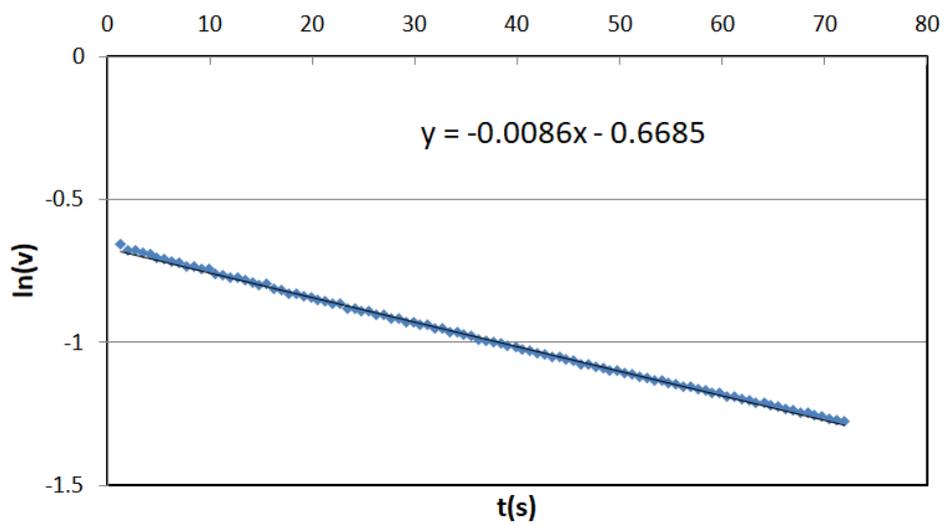

(b)

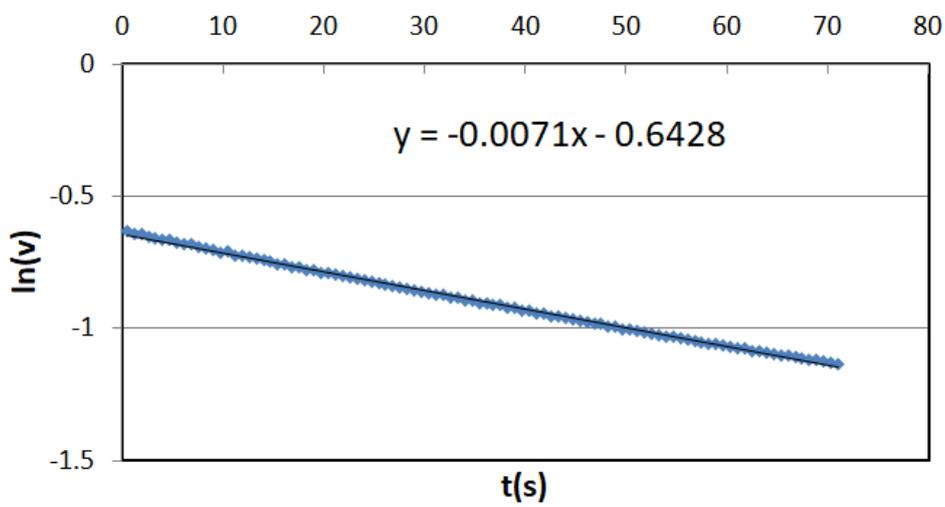

(c)

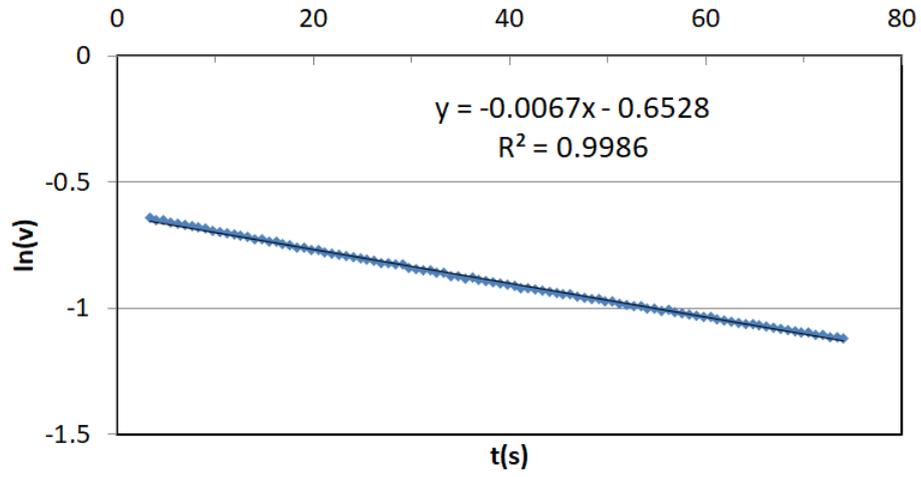

(d)

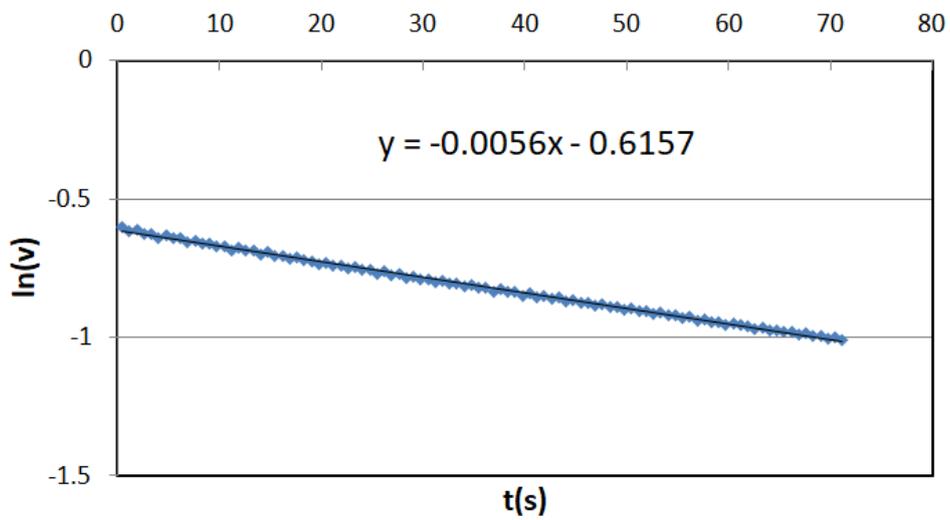

(e)

FIG. 5. Plot of $\ln |v_{max}|$ as a function of time for the simple pendulum motion of the balls of various sizes: (a) 11.5, (b) 18.7, (c) 24.1, (d) 30.0, and (e) 36.0 mm.

The resistance coefficient was obtained from $2m\gamma$ by using Eq. (8). The damping and resistance coefficients measured for the various ball sizes are presented in Table I. As the ball size increased, the resistance coefficient increased. However, the damping coefficient decreased with an increase in the ball size. This implies that when the size of the ball increased, the increase in c was less than that in the mass of the ball.

TABLE I. Measured damping and resistance coefficients for various ball sizes.

| Ball radius (mm) | Mass (g) | Damping coefficient $\gamma$ (1/s) | Resistance coefficient $c$ (kg/s) |
|---|---|---|---|
| 5.75 | 6.36 | 0.0096 | $1.2 \times 10^{-4}$ |
| 9.35 | 8.33 | 0.0086 | $1.4 \times 10^{-4}$ |
| 12.05 | 12.35 | 0.0070 | $1.8 \times 10^{-4}$ |
| 15.00 | 16.28 | 0.0067 | $2.2 \times 10^{-4}$ |
| 18.00 | 23.48 | 0.0056 | $2.6 \times 10^{-4}$ |

A plot of the experimentally observed resistance coefficient as a function of the ball radius is shown in Fig. 6. A linear equation fits the experimental data, except for the smallest ball radius. In order to determine the linear relationship between the resistance coefficient and the ball radius, we assumed $c$ to be given by

$$c = kR^n. \tag{16}$$

Taking the natural logarithm of both sides of this equation yields the following equation:

$$\ln c = \ln k + n \ln R. \quad (17)$$

Therefore, the slope of the plot of $\ln c$ as a function of $\ln R$ is the exponent n in Eq. (16). The value of n obtained from the logarithm plot was 0.975, which is close to 1. This indicated that the resistance coefficient was linearly proportional to the ball radius. The first data point was an outlier to the fitted curve, probably because other friction mechanisms may have come into play for the small ball radius and small friction. Furthermore, the y-intercept of the fitted curve in Fig. 6 is close to zero, which also indicates conformance of the experimental data with the drag equation for small Reynolds numbers.

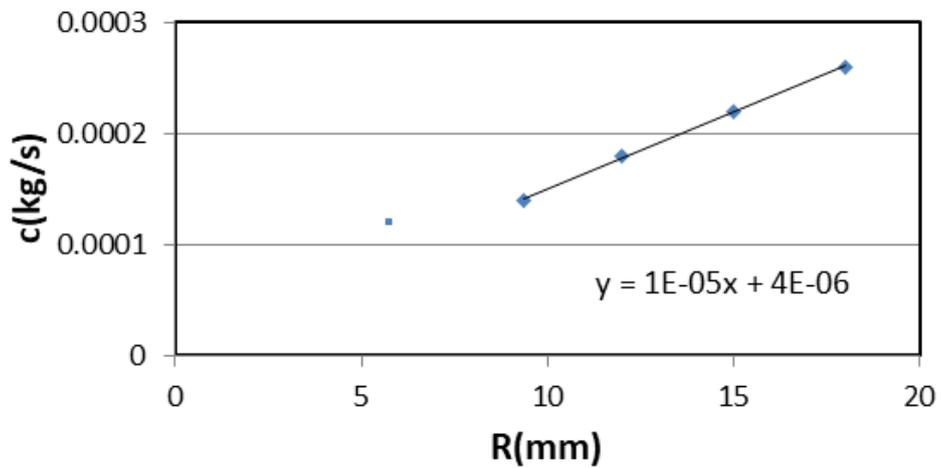

FIG. 6. Plot of the resistance coefficient c vs. the ball radius R.

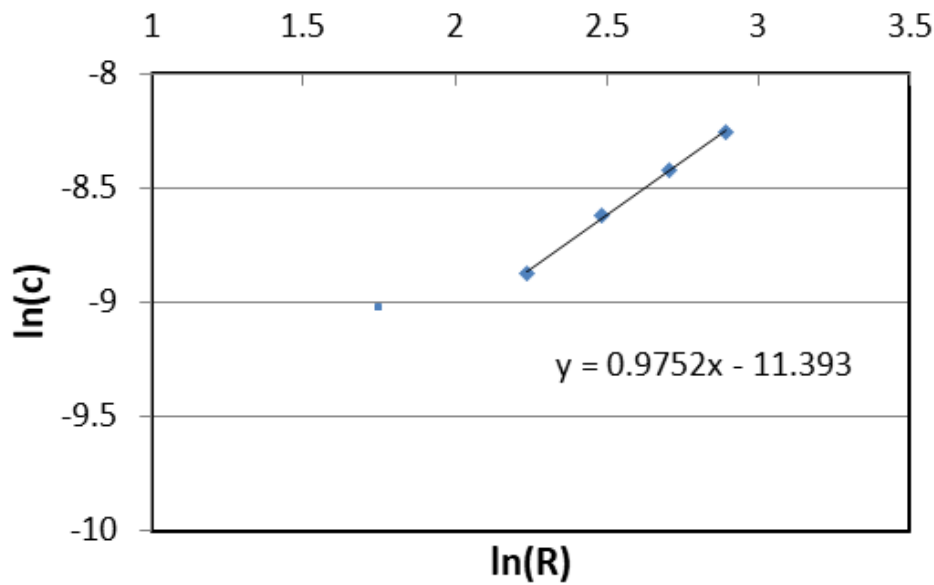

FIG. 7. Plot of $\ln c$ vs. ln R.

## V. CONCLUSIONS

We fabricated a photogate using Arduino and a 3D printer to measure the resistance coefficient of a simple pendulum. Pendulum balls of various sizes were fabricated by inserting a metal ball in 3D-printed plastic shells of various sizes. These balls of various sizes were used to determine the relationship between the resistance coefficient and ball size. The lab-made photogate and Arduino board could measure the time span of the blocking of the laser path by a ball with sufficient accuracy. From the measured time spans, the speed of the balls at their lowest point, which was expected to decrease with the damping coefficient because of air friction, was determined.

The diameters of the balls used in the experiment were 11.5, 18.7, 24.1, 30.0, and 36.0 mm. The experimentally measured damping coefficients for these balls were 0.0096, 0.0086, 0.0070,

0.0067, and 0.0056 1/s, respectively. The resistance coefficients for the balls were obtained by multiplying the damping coefficient by two times the mass of the balls, and the values obtained were $1.2 \times 10^{-4}$, $1.4 \times 10^{-4}$, $1.8 \times 10^{-4}$, $2.2 \times 10^{-4}$, and $2.6 \times 10^{-4}$ kg/s, respectively.

From a plot of the resistance coefficient as a function of the ball radius, the linear relationship between these parameters was obtained and it was observed to be in agreement with the drag equation for small Reynolds numbers. Although the smallest diameter was an outlier to the fitted curve, the logarithmic plot of the resistance coefficients against the ball radius also confirmed that the relationship between the resistance coefficient and the ball radius was linear, in accordance with the drag equation. Thus, measuring the damping and resistance coefficients for pendulum motion was possible using a 3D printer and Arduino. Furthermore, examination of the variation of the resistance coefficient with the ball radius showed that the resistance coefficient was proportional to the ball diameter, in agreement with the drag equation for small Reynolds numbers.

This study shows the possibility of designing physics experiments based on Arduino and a 3D printer for educating students in multi-disciplinary approaches such as those pertaining to mechanics, optics, electronics, and software coding. Furthermore, such experiments cover many aspects of a discipline, such as oscillation, damping, and fluid mechanics. Our results show how multidisciplinary education can be provided to students and educators in science, technology, engineering, and mathematics (STEM).

# REFERENCES

* ygju@knu.ac.kr